%% file: main.tex
\documentclass[conference]{IEEEtran}
\IEEEoverridecommandlockouts

\usepackage{cite}
\usepackage{amsmath,amssymb,amsfonts}
\usepackage{graphicx}
\usepackage{textcomp}
\def\BibTeX{{\rm B\kern-.05em{\sc i\kern-.025em b}\kern-.08em
    T\kern-.1667em\lower.7ex\hbox{E}\kern-.125emX}}

\usepackage{pgf-pie}
\usepackage{xcolor,colortbl}
\usepackage{forloop}
\usepackage{minted}
\usepackage{listings}
\usepackage{algorithm, algpseudocode}
\usepackage{tikz}
\usetikzlibrary{arrows, automata, positioning, bending}
\usetikzlibrary{arrows.meta, positioning}
\usepackage{url}
\usepackage{caption}
\usepackage{subcaption}
\usepackage{pifont}

\begin{document}

\title{BGPFuzz: Automated Configuration Fuzzing of the Border Gateway Protocol
}

\author{
\IEEEauthorblockN{
Chenlu Zhang,
Amirmohammad Pasdar,
Van-Thuan Pham
}
\IEEEauthorblockA{
\textit{School of Computing and Information Systems} \\
\textit{The University of Melbourne} \\
Melbourne, Australia \\
zhanglv0413@gmail.com \\
Amirmohammad.Pasdar@unimelb.edu.au \\
thuan.pham@unimelb.edu.au
}
}

\maketitle

\begin{abstract}
\input{content/abstract}
\end{abstract}


\section{Introduction}
\input{content/introduction}

\section{Background and Related Work}
\input{content/background}

\section{Approach}
This section presents the design and workflow of \textit{BGPFuzz}, a fuzz testing framework for identifying BGP misconfigurations in virtualized networks.

\input{content/approach}

\section{Preliminary Evaluation}
\input{content/result}


\section{Future Work}
\input{content/conclusion}

\bibliographystyle{IEEEtran}
\bibliography{main}

\end{document}

%% file: content/abstract.tex
Telecommunications networks rely on configurations to define routing behavior, especially in the Border Gateway Protocol (BGP), where misconfigurations can lead to severe outages and security breaches—as demonstrated by the 2021 Facebook outage. Unlike existing approaches that rely on synthesis or verification, our work offers a cost-effective method for identifying misconfigurations resulting from BGP's inherent complexity or vendor-specific implementations. We present \textit{BGPFuzz}, a structure-aware and stateful fuzzing framework that systematically mutates BGP configurations and evaluates their effects in virtualized network. Without requiring predefined correctness properties as in static analysis, BGPFuzz detects anomalies through runtime oracles that capture practical symptoms such as session resets, blackholing, and traffic redirection. Our experiments show that BGPFuzz can reliably reproduce and detect known failures, including max-prefix violations and sub-prefix hijacks. 

%% file: content/introduction.tex
\begin{figure}[tbp]
\centerline{\includegraphics[width=0.5\textwidth]{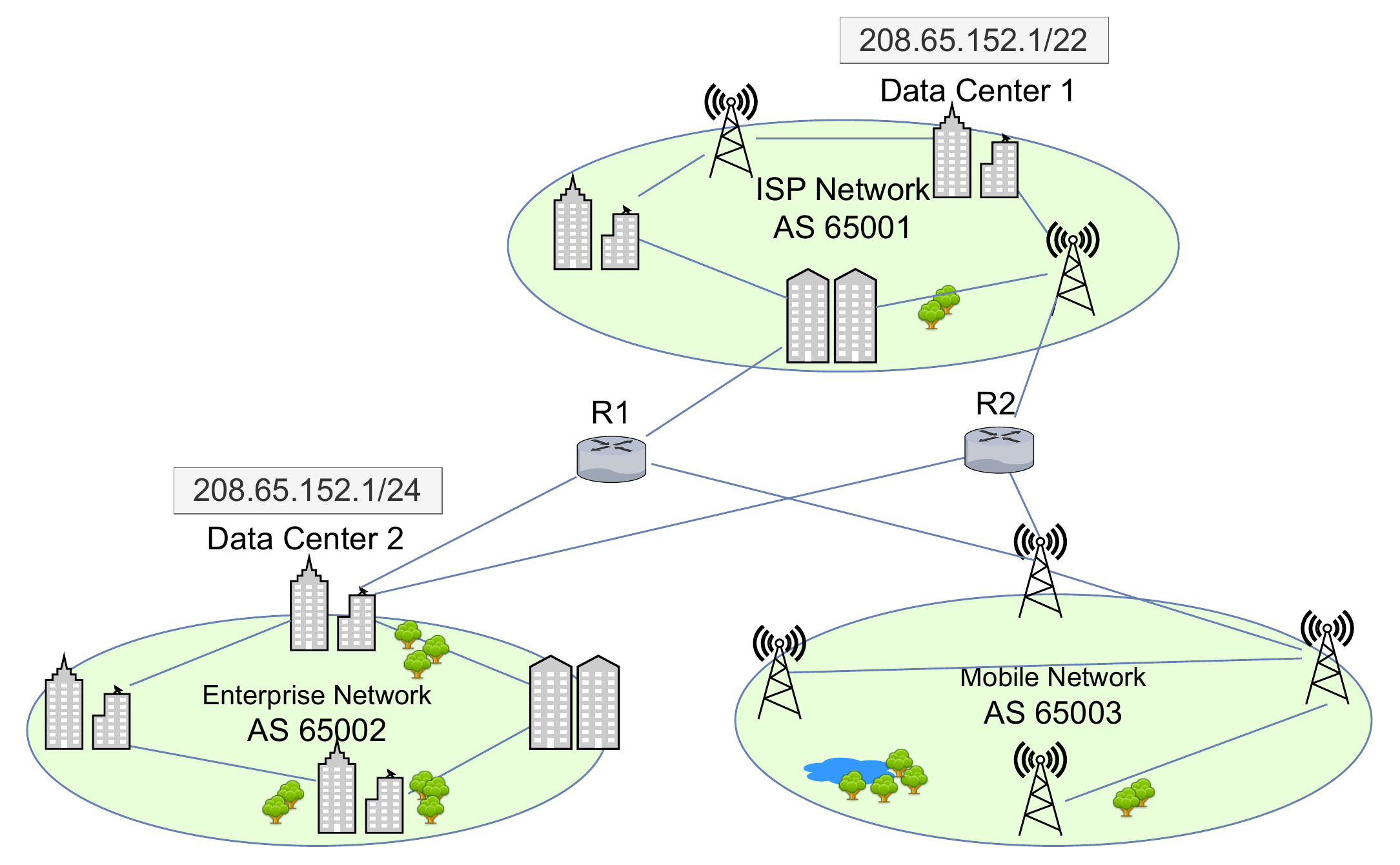}}
\caption{An example telecommunications network illustrating connections between different Autonomous Systems (ASes).}
\label{fig:sub_prefix_topo}
\end{figure}

Telecommunication networks rely on carefully designed configurations that define routing behavior, security policies, and data flow between interconnected systems. These configurations often include protocols such as Open Shortest Path First (OSPF), Intermediate System to Intermediate
System (IS-IS), and the Border Gateway Protocol (BGP), which determines interdomain routing between Autonomous Systems (ASes)~\cite{DBLP:journals/rfc/rfc1772}. Network configurations are initially created during deployment and continually updated to accommodate infrastructure changes, policy revisions, or application demands~\cite{DBLP:conf/sigcomm/LiuWZCWXZM018}. Human network operators must translate high-level intent—such as the example in Fig.\ref{fig:sub_prefix_topo}, where AS 65001 prefers a specific neighbor (R1) over another (R2)—into low-level, device-specific configurations. However, the scale and heterogeneity of modern networks introduce significant complexity, making this process highly error-prone\cite{DBLP:conf/sigcomm/SungTWZ16}.

Misconfigurations in routing protocols, particularly BGP, are a leading cause of network disruptions. For instance, the 2021 Facebook outage was caused by cascading router updates following a misapplied configuration, resulting in a global six-hour service disruption~\cite{FireMon2021}.Historically, BGP misconfigurations have caused incidents such as the 2008 global YouTube outage~\cite{youtubehijack2008}, cryptocurrency redirection attacks~\cite{myetherwallet2018}, and cloud service disruptions~\cite{cloudflareblackhole, amazonroute53, DBLP:conf/pam/HiranCG13, qrator2022}. Reports from industry also highlight that more than 60\% of network incidents stem from misconfigurations~\cite{DBLP:conf/sigcomm/LiuWZCWXZM018}. Hence, the need for automated and effective validation techniques is critical for improving the robustness of backbone networks.

Existing approaches to mitigating network misconfigurations generally fall into two categories: synthesis and verification. Configuration synthesis automatically generates device configurations from human-defined high-level intent to reduce manual effort and human error~\cite{DBLP:conf/nsdi/El-HassanyTVV18, DBLP:conf/sigcomm/SchneiderBV21, DBLP:journals/comsur/TroisFBM16}. However, this approach requires that the high-level intent be robust and well-defined in real-world scenarios—an assumption that is not always guaranteed. Therefore, an additional step of proof or testing is often necessary. In contrast, verification ensures that a given configuration satisfies specified safety properties, either statically~\cite{DBLP:journals/comcom/GritzalisSG99, DBLP:conf/sigcomm/SteffenGTVV20} or dynamically at runtime~\cite{DBLP:conf/ccs/MengPRS23, DBLP:conf/nsdi/SinghaMBKMV24}. 

Runtime verification is useful for large-scale and stateful systems, where static methods often fail to handle scale and dynamic behavior~\cite{DBLP:journals/comcom/GritzalisSG99}. For example, existing runtime verification for BGP configurations contributes to detecting flaws—violations of the RFC specification~\cite{DBLP:journals/rfc/rfc4271}—in BGP implementations by different vendors~\cite{DBLP:conf/nsdi/SinghaMBKMV24}. In contrast, our work targets vulnerabilities that are inherent to BGP itself, such as the over-trust of peers. These issues can occur even when BGP is implemented correctly according to the RFC.

To identify BGP misconfigurations, we adopt a fuzzing-based approach that first synthesizes configurations and then applies them to a simulated network for runtime verification. This offers a promising direction toward scalable and robust network testing. Existing state-of-the-art network protocol fuzzers consider network state by inferring it from packet content~\cite{DBLP:conf/icst/PhamBR20, DBLP:conf/ipccc/CaiYLX23}. However, these approaches are packet-driven~\cite{DBLP:conf/ipccc/CaiYLX23}. They inject malformed packets into the network, which may not reflect errors that naturally arise from configuration updates. In contrast, our work focuses on misconfigurations made by network operators, where the resulting packet flows are realistically produced by configuration changes rather than deliberate attacks—highlighting a distinct and underexplored threat model.

In this work, we propose \textit{BGPFuzz}, an automated and expansible testing framework to identify misconfigurations in BGP configurations. It leverages structure-aware test case generation, runtime network state-driven testing, and multi-anomaly detection—enabled by multiple test oracles—to automatically generate and validate test cases.
Unlike existing protocol fuzzers~\cite{DBLP:conf/icst/PhamBR20, DBLP:journals/itiis/WangZL13}, BGPFuzz is tailored for the semantics of BGP configurations and addresses several challenges, including (1) centralized vs. decentralized test coordination~\cite{DBLP:conf/pts/UlrichK99, DBLP:conf/nsdi/GuptaVV08}, (2) stateful protocol behavior~\cite{DBLP:conf/ccs/MengPRS23}, (3) balancing input validity and generation efficiency~\cite{DBLP:conf/icse/Pham23}, and (4) detecting non-obvious yet disruptive behaviors, including route leaks and session flaps~\cite{DBLP:journals/tse/ManesHHCESW21}.

To address these challenges, we make four key design decisions. First, we adopt a centralized testing architecture based on simulated environments, as we aim to avoid disruption to real networks~\cite{DBLP:journals/cn/GomezKCS23}. Second, BGPFuzz maintains state awareness by observing runtime information of the network and adapting input generation accordingly. Third, it incorporates structure awareness through grammar-based test case generation, significantly reducing invalid test cases. Finally, it integrates multi-anomaly detection to capture diverse failure modes including max-prefix violations~\cite{Optus2023Outagebgpf}, sub-prefix hijacks~\cite{youtubehijack2008}, and session instabilities~\cite{DBLP:journals/jsac/SriramMBKK06}.


\noindent In summary, this paper makes the following contributions:
\begin{itemize}
\item We design and implement \textit{BGPFuzz}, the first stateful and structure-aware configuration fuzzing framework for BGP, to the best of our knowledge.
\item We incorporate complementary runtime oracles to detect real-world BGP anomalies with high accuracy.
\item We conduct preliminary evaluations demonstrating that BGPFuzz outperforms random fuzzing in terms of bug discovery rate, input validity, and test stability.
\end{itemize}

%% file: content/background.tex

Recent network incidents~\cite{youtubehijack2008, myetherwallet2018, amazonroute53, cloudflareblackhole, DBLP:conf/pam/HiranCG13, qrator2022} highlight a core BGP vulnerability: its default trust in neighboring routers, due to the lack of built-in validation~\cite{DBLP:journals/rfc/rfc4272}. BGP uses the longest prefix match rule, favoring the most specific route without verifying if the announcer is authorized to advertise the prefix. This allows unauthorized announcements—whether malicious or accidental—to hijack traffic, often without disrupting apparent connectivity. For instance, in Fig.~\ref{fig:sub_prefix_topo}, a more specific prefix advertised by Data Center 2 causes traffic to be misrouted away from Data Center 1.

BGP lacks resilience to many threats. It has no built-in checks for prefix ownership, peer authenticity, or path integrity~\cite{DBLP:journals/rfc/rfc4272}, and omits encryption, replay protection, and message authentication—leaving it vulnerable to spoofing, man-in-the-middle attacks, and prefix flooding. Even normal changes like link failures or policy updates can trigger route shifts that mimic attacks, complicating detection. Improving BGP is difficult, as protocol or vendor changes require global router updates—comparable to replacing IPv4. To avoid this, researchers proposed a decentralized patch layer that logs local BGP decisions and uses global consensus to ensure consistent routing~\cite{DBLP:conf/nsdi/JohnKKAV08}. While effective without altering BGP, this still requires broad adoption and further refinement.

Several approaches aim to prevent or detect BGP misconfigurations. Configuration synthesis tools like NetComplete~\cite{DBLP:conf/nsdi/El-HassanyTVV18} and Propane~\cite{DBLP:conf/sigcomm/BeckettMMPW16} generate router configurations from high-level intent by modeling routing constraints and automating translation. However, they do not explicitly ensure safety when applied to dynamic or heterogeneous network environments. Snowcap~\cite{DBLP:conf/sigcomm/SchneiderBV21} improves upon synthesis by preserving safety during incremental updates, but still requires manual intervention for intermediate configurations. Configuration verification techniques span static and runtime methods. Formal approaches verify protocol correctness~\cite{DBLP:journals/comcom/GritzalisSG99}, while probabilistic ones like NetDice~\cite{DBLP:conf/sigcomm/SteffenGTVV20} assess fault tolerance under link failures. Tools like MESSI~\cite{DBLP:conf/nsdi/SinghaMBKMV24} and RPFuzzer~\cite{DBLP:journals/itiis/WangZL13} use symbolic execution and feedback-driven testing for runtime verification. However, they focus on implementation bugs, not misconfiguration-induced failures.

All the methods discussed above begin with simulation or emulation environments, which offer controlled settings for reproducible experiments but often lack realism. In contrast, testbeds provide more realistic conditions but are costly and inaccessible to many researchers~\cite{DBLP:journals/cn/GomezKCS23}. Among various simulators—such as NS-3, OMNeT++, and GNS3—GNS3~\cite{GNS3} stands out for supporting real router operating systems and full configuration syntax. This enables precise modeling of inter-domain BGP behavior and the impact of misconfigured settings under controlled yet realistic conditions.

Fuzzing is a cost-effective, automated technique for exploring misconfiguration impacts in simulated or real networks. It executes programs with generated inputs—often malformed—to trigger bugs and expose vulnerabilities~\cite{DBLP:journals/tse/ManesHHCESW21}. In protocol fuzzing, mutation-based methods often generate invalid cases, reducing efficiency~\cite{DBLP:conf/ndss/MengMBR24}. Improving test input quality enhances effectiveness. Tools like Superion~\cite{DBLP:conf/icse/Wang0WL19} and AFLSmart++~\cite{DBLP:conf/icse/Pham23} boost input validity by respecting structure, making them well-suited for complex configurations like BGP.

Protocol fuzzing must handle stateful behavior using runtime feedback—such as message responses or protocol states—to guide test generation. Techniques include response-code tracking~\cite{DBLP:conf/icst/PhamBR20}, memory observation, and reinforcement learning~\cite{DBLP:conf/ccs/MengPRS23}. Oracles enhance fuzzing by detecting anomalies via system logs~\cite{DBLP:conf/uss/ZhangBLD0024}, metamorphic relations~\cite{DBLP:conf/icse/PanCMP24}, or abnormal resource usage~\cite{DBLP:conf/icsai/LiZC14}. Despite growing interest, no prior work targets BGP configuration testing. Existing fuzzers focus on traffic or implementations, overlooking the structure and semantics of router configs. \textit{BGPFuzz} addresses this gap as the first structure-aware, stateful framework for validating BGP configurations, combining grammar-based generation, runtime guidance, and multi-oracle detection to uncover misconfiguration-induced anomalies.

%% file: content/approach.tex
\subsection{Framework Overview}
\label{subsec:framework-overview}


\begin{figure}[tbp]
\centerline{\includegraphics[width=0.5\textwidth]{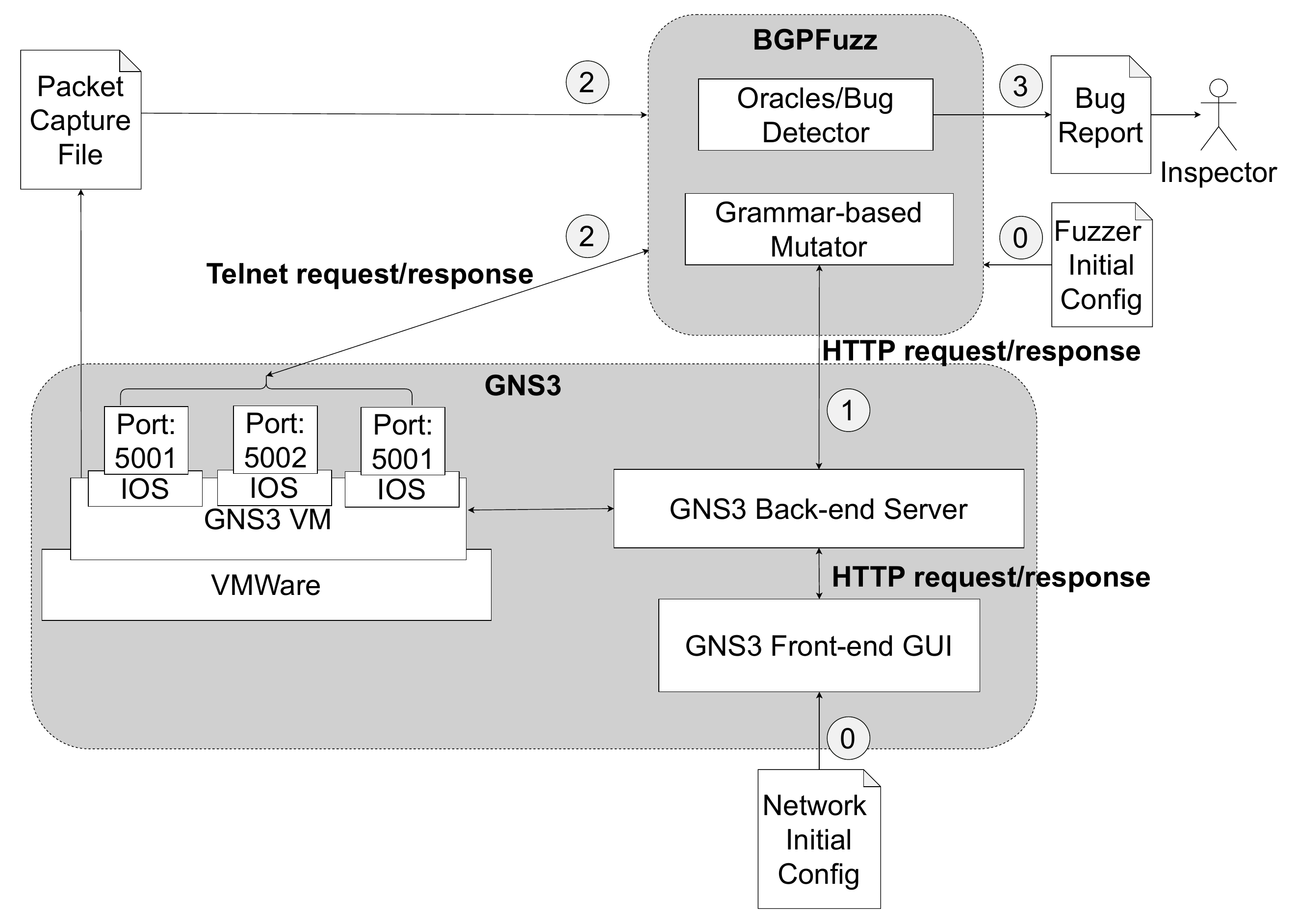}}
\caption{The workflow of our approach to identify BGP vulnerabilities in a given network topology.}
\label{fig:bgpfuzz-workflow}
\end{figure}

    The overall architecture and execution pipeline of \textit{BGPFuzz} is shown in Fig.~\ref{fig:bgpfuzz-workflow}. The framework consists of four sequential steps: Step-0 handles initialization, while Steps-1 to 3 form the iterative core fuzz testing process.

\textbf{Step-0: Topology Initialization and Target Selection. }
The process begins with manual setup. The tester builds a BGP topology in the network simulator GNS3 using Cisco Internetwork Operating System (IOS) router images, connects nodes to form BGP sessions, and applies valid initial configurations—including IP addressing, interface setup, and BGP neighbor definitions—to ensure stable protocol convergence before mutation. A target router node is then selected for testing, along with a specific network interface for traffic capture. This step sets the focus of the test. It is not automated. Experts choose the topology and plan realistic scenarios.  For realism, topologies can be selected from the Topology Zoo~\cite{DBLP:journals/jsac/KnightNFBR11}, a collection of real-world network topologies.

\textbf{Step-1: Configuration Mutation and Deployment. }
\textit{BGPFuzz} extracts the current configuration of the selected node as a seed and generates mutated versions using structure-aware (i.e., grammar-aware) rules based on the grammar of BGP configuration files. It constructs a grammar derivation tree~\cite{DBLP:books/fuzzingbook2024} to guide this process. Listing~\ref{lst:bgp-ip-highlight} shows an example BGP configuration, where the highlighted parts represent fields subject to mutation. These mutated configurations are syntactically valid and adhere to the grammar of BGP directives, thereby reducing invalid test cases and improving testing efficiency. The mutated configuration is deployed to the router via GNS3's REST API, and the node is refreshed to apply the changes.

\lstdefinestyle{bgpstyle}{
    basicstyle=\ttfamily\small,
    keywordstyle=\color{black},
    commentstyle=\color{gray},
    stringstyle=\color{blue},
    columns=fullflexible,
    keepspaces=true,
    morekeywords={router, bgp, address-family, neighbor, network, activate, exit-address-family, remote-as},
    moredelim=**[is][\color{red}]{@}{@} 
}

\begin{figure}[hbtp]
    \centering
    \begin{minipage}{0.45\textwidth}
    \lstset{
      backgroundcolor=\color{gray!10},
      basicstyle=\scriptsize\ttfamily,
      breaklines=true,
      prebreak=\raisebox{0ex}[0ex][0ex]{\ensuremath{\hookleftarrow}},
      frame=single,
      escapeinside={(*@}{@*)},
      captionpos=b,
      commentstyle=\color{black},
      morecomment=[l]{//},
      morecomment=[s]{/*}{*/}
    }
    \begin{lstlisting}[style=bgpstyle, caption={Example BGP Configuration Highlighting Mutatable Fields}, label={lst:bgp-ip-highlight}]
router bgp 45000
 router-id @172.17.1.99@
 bgp log-neighbor-changes
 neighbor @192.168.1.2@ remote-as @40000@
 neighbor @192.168.3.2@ remote-as @50000@
    \end{lstlisting}
    \end{minipage}
\end{figure}

\textbf{Step-2: Monitoring and Feedback Collection. }
After applying the mutated configuration, BGP sessions are reestablished and the network stabilizes. \textit{BGPFuzz} then captures traffic on the selected interface and optionally queries internal router state (e.g., via packet captures and Telnet responses) to collect network traffic and routing table data. These runtime observations provide feedback for guiding the next round of mutations, enabling state-aware test generation and improving fuzzing efficiency.

\textbf{Step-3: Oracle Evaluation and Recovery. }
The captured network data (i.e., the packet capture file and Telnet responses from Step-2 in Fig.~\ref{fig:bgpfuzz-workflow}) is analyzed by a set of oracles that detect symptoms of misconfiguration or abnormal BGP behavior—such as session flaps, unreachable routes, or unexpected redirections. If an anomaly is detected, the corresponding configuration is saved for further inspection, and the system restores all nodes to their initial state to ensure a clean environment for the next iteration. Otherwise, \textit{BGPFuzz} proceeds directly to Step-1, continuing test case generation and deployment without recovery.

This modular workflow enables repeatable and guided exploration of BGP misconfiguration scenarios with minimal manual intervention per iteration.

\subsection{Input Quality Improvement}
\label{subsec:input-quality}

To reduce the cost of invalid test cases and improve bug discovery, \textit{BGPFuzz} adopts both structure-aware and state-aware input generation strategies.

\subsubsection{Structure-Aware Mutation Strategy}

To improve input validity during fuzz testing, \textit{BGPFuzz} uses a structure-aware mutation strategy. Traditional mutation-based fuzzers rely on low-level operations like bit flips or deletions, which often ignore input structure and generate invalid configurations rejected by parsers, reducing fuzzing efficiency~\cite{DBLP:journals/tse/ManesHHCESW21}. \textit{BGPFuzz} addresses this by applying mutations guided by the formal grammar of BGP configuration directives, producing syntactically valid and semantically meaningful updates. This increases the likelihood of test cases being accepted and executed. Beyond grammar conformance, \textit{BGPFuzz} prioritizes high-impact fields, as BGP misconfigurations can lead to harmful but protocol-compliant behaviors. Based on RFC 4272 and incident analyses, about 18\% of impactful failures (e.g., session resets, hijacks) are triggerable via configuration, while 53\% require raw packet crafting. \textit{BGPFuzz} focuses on the configuration-accessible subset, using grammar-aware and feedback-driven fuzzing to uncover realistic, operator-induced BGP faults.



\subsubsection{State-Aware Input Generation}

\begin{figure}[tb]
    \centering
    \resizebox{0.30\textwidth}{!}{%
    \begin{tikzpicture}[
        ->, 
        >=stealth',
        node distance=3.5cm and 2.8cm,
        semithick,
        auto,
        state/.style={circle, draw, minimum size=1.4cm, font=\small},
        scale=0.9, transform shape
    ]
        \node[state] (s0) {$S_0$};
        \node[state] (s1) [right of=s0] {$S_1$};
        \node[state] (s2) [right of=s1] {$S_2$};

        \path (s0) edge[loop above] node{$E_1$/$E_2$/$E_3$} (s0);
        \path (s1) edge[loop above] node{$E_1$/$E_2$/$E_3$/$E_4$} (s1);

        \path (s0) edge node{$E_1$/$E_2$/$E_3$} (s1);
        \path (s1) edge node{$E_5$} (s2);

        \path (s0) edge[bend left=70, above] node{$E_5$} (s2);

        \path (s2) edge[bend right=-70] node[below] {$E_6$} (s0);

    \end{tikzpicture}%
    }
    \caption{BGPFuzz Runtime State Machine. States: $S_0$ = Normal Run, $S_1$ = Intermediate (No Error), $S_2$ = Error Detected. Transitions are labeled $E_j$.}
    \label{fig:bgpfuzzfsm}
\end{figure}
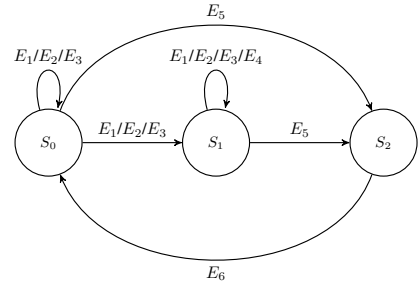

A key feature of \textit{BGPFuzz} is its greybox feedback mechanism, which gathers runtime observations—such as BGP table convergence, prefix announcements, and traffic anomalies—to guide test case generation. This feedback allows the fuzzer to track the network's internal state and make informed mutation decisions, in contrast to random or blackbox fuzzers. \textit{BGPFuzz} models runtime behavior using a state machine with states $S_i$ and transitions $E_j$ (Figure~\ref{fig:bgpfuzzfsm}). States include $S_0$: normal operation, $S_1$: post-mutation with no error, and $S_2$: anomaly detected. Transitions are driven by events: $E_{1\text{--}3}$ (configuration changes), $E_4$ (prefix announcement), $E_5$ (oracle-triggered error), and $E_6$ (recovery). This model enables feedback-guided fuzzing by prioritizing mutations that induce meaningful state shifts. Runtime feedback enables smarter input generation. For instance, the mutator reproduces a sub-prefix hijack scenario by extracting an existing prefix:

\begin{lstlisting}[style=bgpstyle]
*> 208.65.152.0/22 0.0.0.0 0 32768 i
\end{lstlisting}

and synthesizes a more-specific announcement:

\begin{lstlisting}[style=bgpstyle]
network 208.65.152.0 mask 255.255.255.0
\end{lstlisting}

To ensure reachability, it inserts a static route:

\begin{lstlisting}[style=bgpstyle]
ip route 208.65.152.1 255.255.255.0
network 208.65.152.0 mask 255.255.255.0
\end{lstlisting}

Without feedback, test cases may announce unreachable prefixes. By using runtime BGP state, \textit{BGPFuzz} generates valid, high-impact inputs that improve fuzzing effectiveness.

\subsection{Bug Detector Design}
\label{subsec:bug-detector}

\textit{BGPFuzz} detects anomalies using automated oracles based on common BGP failures such as session resets, blackholing, and prefix hijacking, as seen in real-world incidents~\cite{youtubehijack2008, amazonroute53, myetherwallet2018}. Packet-level oracles inspect for BGP \textit{NOTIFICATION} messages (indicating session drops) and ICMP \textit{Destination Unreachable} responses, signaling blackholed prefixes. To uncover subtler issues, metamorphic-relation oracles track routing changes—e.g., unusual AS paths, misrouted traffic delays, or overlapping sub-prefixes. For instance, announcing \texttt{208.65.153.0/24} alongside \texttt{208.65.152.0/22} may suggest a hijack or misconfiguration. All oracles run in parallel, logging anomalies and archiving configurations. This multi-oracle strategy enhances failure coverage and supports robust post-analysis across diverse fault types.

%% file: content/result.tex
\textbf{Implementation.} \textit{BGPFuzz} is implemented in Python 3.12 atop the \texttt{fuzzingbook}~\cite{DBLP:books/fuzzingbook2024} framework. It runs in GNS3 2.2.49 using Cisco 7200 IOS images within VMware Workstation Pro 16. We make our implementation available at \emph{\url{https://github.com/MelbourneFuzzingHub/bgpfuzz}}.

\textbf{Targets \& Baseline.} We evaluate \textit{BGPFuzz} on known issues: invalid configuration (Bug-01), max-prefix violation (Bug-02)~\cite{Optus2023Outagebgpf}, and sub-prefix hijack (Bug-03)~\cite{youtubehijack2008}. We compare \textit{BGPFuzz} with a black-box random baseline lacking grammar or state-awareness, as no BGP configuration-specific fuzzers exist for direct comparison.

\textbf{Setup.} To ensure realism, we use topologies from Topology Zoo~\cite{DBLP:journals/jsac/KnightNFBR11}, grouped as Tiny (5–15 nodes), Small (20–50), and Large (80+). Hardware limits (4-core CPU, 16 GB RAM) restrict testing to Tiny topologies. Larger ones exceeded local and cloud capabilities due to GNS3 GUI constraints. Still, most BGP anomalies emerge in local neighborhoods, making Tiny topologies sufficient.

\begin{table}[htbp]
\centering
\caption{Comparison of Fuzzing Strategies on Bug Detection (10-min run, avg over 10 trials)}
\label{tab:bgpfuzz-bug-table}
\renewcommand{\arraystretch}{1.2}
\resizebox{0.5\textwidth}{!}{%
\begin{tabular}{|l|p{3.2cm}|c|c|}
\hline
\textbf{Bug ID} & \textbf{Bug Type} & \textbf{Random} & \textbf{BGPFuzz} \\
\hline
Bug-01 & Invalid config            & \textbf{\ding{51}}     & -- \\
Bug-02 & Max-prefix violation         & --     & \textbf{\ding{51}} \\
Bug-03 & Sub-prefix hijack         & --     & \textbf{\ding{51}} \\
\hline
\end{tabular}
}
\end{table}

As shown in Table~\ref{tab:bgpfuzz-bug-table}, \textit{BGPFuzz} detects two major misconfiguration-induced failures (Bug-02, Bug-03). Its grammar-based approach avoids generating invalid configurations, so it does not reproduce Bug-01—unsurprisingly the only bug found by black-box random fuzzing. To trigger \textbf{Bug-02} (max-prefix violations~\cite{Optus2023Outagebgpf}), \textit{BGPFuzz} sets a low \texttt{maximum-prefix} limit. Exceeding it causes session resets and \texttt{NOTIFICATION} messages. This \textit{stateless bug} triggers reliably with shallow fuzzing. For \textbf{Bug-03} (sub-prefix hijacks~\cite{youtubehijack2008}), a more specific prefix is injected, exploiting BGP’s longest-prefix match to redirect traffic. This results in blackholing (via ICMP unreachable) or interception (via traceroute shifts). These \textit{stateful failures} highlight \textit{BGPFuzz}'s runtime-aware detection of propagation-based anomalies.


%% file: content/conclusion.tex
Future work includes scaling to larger topologies, deploying in real-world testbeds—ideally in collaboration with Internet service providers—detecting zero-day misconfigurations, and assessing bug severity to prioritize critical cases.